\documentclass[%
 reprint,
 amsmath,amssymb,
 aps,
]{revtex4-2}

\usepackage{graphicx}
\usepackage{dcolumn}
\usepackage{bm}
\usepackage{physics}
\usepackage{caption}
\usepackage{subcaption}
\usepackage{float}
\usepackage{hyperref}
\hypersetup{
    colorlinks=true,
    linkcolor=red,
    filecolor=magenta,
    urlcolor=cyan,
    citecolor=blue
}

\begin{document}

\preprint{APS/123-QED}

\title{The Zeno and anti-Zeno effects: studying modified decay rates for spin-boson models with both strong and weak system-environment couplings}
\thanks{It is a paper on a possible extension of the work done in Ref.~\cite{chaudhry2017quantum}.}%

\author{Irfan Javed}
 \email{irfan.devaj@outlook.com}
\author{Mohsin Raza}
 \email{mohsinrazaonline@gmail.com}
 
\affiliation{%
 School of Science and Engineering, Lahore University of Management Sciences (LUMS), Opposite Sector U, D.H.A., Lahore 54792, Pakistan
}%

\date{\today}

\begin{abstract}
In this paper, we look into what happens to a quantum system under repeated measurements if system evolution is removed before each measurement is performed. Beginning with investigating a single two-level system coupled to two independent baths of harmonic oscillators, we move to replacing it with a large collection of such systems, thereby invoking the large spin-boson model. Whereas each of our two-level systems interacts strongly with one of the aforementioned baths, it interacts weakly with the other. A polaron transformation is used to make it possible for the problem in the strong coupling regime to be treated with perturbation theory. We find that the case involving a single two-level system exhibits qualitative and quantitative differences from the case involving a collection of them; however, the general effects of strong and weak couplings turn out to be the same as those in the presence of system evolution, something which allows us to establish that system evolution has no practical bearing on any of these effects.
\end{abstract}

\maketitle


\section{\label{sec. 1}Introduction}
We start with the paradigmatic spin-boson model~\cite{leggett1987dynamics}, but we consider the presence of both strong and weak system-environment couplings. The Hamiltonian describing the problem is the following:
\begin{equation}
\begin{aligned}[b]
H_{L}^{(0)} =\:&\frac{\epsilon}{2}\sigma_{z}+\frac{\Delta}{2}\sigma_{x}+\sum_{k}\omega_{k}b_{k}^{\dagger}b_{k}+\sum_{k}\alpha_{k}c_{k}^{\dagger}c_{k}\\
&+\sigma_{z}\sum_{k}\left(g_{k}^{*}b_{k}+g_{k}b_{k}^{\dagger}\right)+\sigma_{x}\sum_{k}\left(f_{k}^{*}c_{k}+f_{k}c_{k}^{\dagger}\right).
\end{aligned}
\label{eq. 1}
\end{equation}
Here, $\frac{\epsilon}{2}\sigma_{z}+\frac{\Delta}{2}\sigma_{x}$ is the system Hamiltonian, $\sum_{k}\omega_{k}b_{k}^{\dagger}b_{k}+\sum_{k}\alpha_{k}c_{k}^{\dagger}c_{k}$ is the environment Hamiltonian, and $\sigma_{z}\sum_{k}\left(g_{k}^{*}b_{k}+g_{k}b_{k}^{\dagger}\right)+\sigma_{x}\sum_{k}\left(f_{k}^{*}c_{k}+f_{k}c_{k}^{\dagger}\right)$ gives the system-environment interaction. $\epsilon$ characterizes the energies of the system energy eigenstates, $\Delta$ is the tunneling amplitude, and $\omega_{k}$ and $\alpha_{k}$ are the frequencies of harmonic oscillators in the two harmonic oscillator baths interacting with the system. $b_{k}/b_{k}^{\dagger}$ and $c_{k}/c_{k}^{\dagger}$ are the annihilation/creation operators of the first and second baths, respectively, $\sigma_{x}$ and $\sigma_{z}$ are the standard Pauli operators, and we set $\hbar$ equal to $1$ throughout. Superscript $(0)$ denotes the first version of our Hamiltonian, subscript $L$ denotes the lab frame, and henceforth, we use the following definitions: $H_{S1} = \frac{\epsilon}{2}\sigma_{z}$, $H_{S2} = \frac{\Delta}{2}\sigma_{x}$, $H_{B1} = \sum_{k}\omega_{k}b_{k}^{\dagger}b_{k}$, $H_{B2} = \sum_{k}\alpha_{k}c_{k}^{\dagger}c_{k}$, $V_{C1} = \sigma_{z}\sum_{k}\left(g_{k}^{*}b_{k}+g_{k}b_{k}^{\dagger}\right)$, and $V_{C2} = \sigma_{x}\sum_{k}\left(f_{k}^{*}c_{k}+f_{k}c_{k}^{\dagger}\right)$.

As per our Hamiltonian, a spin-$\frac{1}{2}$ (two-level) system interacts with an environment comprised of two independent baths of harmonic oscillators. Whereas one bath interacts with the $z$ component of the spin of the system, the other interacts with the $x$ component. All through this paper, we assume that interaction, or coupling, with the $z$ component is strong and that interaction, or coupling, with the $x$ component is weak, something which implies that $\abs{g_{k}}>\abs{f_{k}}$ and $\abs{f_{k}}<1$.

If the system is prepared in the state $\ket{\uparrow}$, then it evolves both due to the tunneling term $\frac{\Delta}{2}\sigma_{x}$ and due to the system-environment couplings. However, interested in changes in the system state stemming from the system-environment interactions only, we remove the evolution due to our system Hamiltonian, $H_{S1}+H_{S2}$, before performing any measurement to check the state of the system~\cite{matsuzaki2010quantum}. The effective decay rate obtained as a result is what we call the modified decay rate, and we investigate how it depends on the strengths of the strong and weak system-environment couplings.

We apply this very treatment to the problem generalizing our spin-boson model to $N_{S}$ two-level systems, all of which interact with the aforementioned environment of harmonic oscillator baths in the same way as the single two-level system described above; the only difference is that we prepare our system in a coherent spin state rather than $\ket{\uparrow}$ this time. Also known as the large spin-boson model~\cite{chaudhry2013role}, this problem is described by the following Hamiltonian:
\begin{equation}
\begin{aligned}[b]
H_{L}^{(1)} =\:&\epsilon J_{z}+\Delta J_{x}+\sum_{k}\omega_{k}b_{k}^{\dagger}b_{k}+\sum_{k}\alpha_{k}c_{k}^{\dagger}c_{k}\\
&+2J_{z}\sum_{k}\left(g_{k}^{*}b_{k}+g_{k}b_{k}^{\dagger}\right)\\
&+2J_{x}\sum_{k}\left(f_{k}^{*}c_{k}+f_{k}c_{k}^{\dagger}\right),
\end{aligned}
\label{eq. 2}
\end{equation}
where $J_{x}$, $J_{y}$, and $J_{z}$ are the usual angular momentum operators obeying the commutation relations $[J_{k}, J_{l}] = \iota\epsilon_{klm}J_{m}$; superscript $(1)$ denotes the second version of our Hamiltonian; and all other symbols have the meanings ascribed to them in the paragraph following Eq.~(\ref{eq. 1}). Moreover, for this problem, we use the following definitions henceforth: $H'_{S1} = \epsilon J_{z}$, $H'_{S2} = \Delta J_{x}$, $H'_{B1} = \sum_{k}\omega_{k}b_{k}^{\dagger}b_{k}$, $H'_{B2} = \sum_{k}\alpha_{k}c_{k}^{\dagger}c_{k}$, $V'_{C1} = J_{z}\sum_{k}\left(g_{k}^{*}b_{k}+g_{k}b_{k}^{\dagger}\right)$, and $V'_{C2} = J_{x}\sum_{k}\left(f_{k}^{*}c_{k}+f_{k}c_{k}^{\dagger}\right)$.

\section{\label{sec. 2}Results}
\subsection{\label{sec. 2a}Spin-boson model}
\subsubsection{\label{sec. 2aa}System density matrix}
In this case, our Hamiltonian is $H_{L}^{(0)}$ (Eq.~(\ref{eq. 1})), and we use a polaron transformation defined by $U_{P} = e^{\chi\sigma_{z}/2}$, where $\chi = \sum_{k}\left(\frac{2g_{k}}{\omega_{k}}b_{k}^{\dagger}-\frac{2g_{k}^{*}}{\omega_{k}}b_{k}\right)$, to transform it~\cite{silbey1984variational}. In the polaron frame then, the Hamiltonian is
\begin{equation}
\begin{aligned}[b]
H^{(0)} =\:&\frac{\epsilon}{2}\sigma_{z}+\sum_{k}\omega_{k}b_{k}^{\dagger}b_{k}+\sum_{k}\alpha_{k}c_{k}^{\dagger}c_{k}\\
&+\left[\frac{\Delta}{2}+\sum_{k}\left(f_{k}^{*}c_{k}+f_{k}c_{k}^{\dagger}\right)\right]\left(\sigma_{+}e^{\chi}+\sigma_{-}e^{-\chi}\right).
\end{aligned}
\label{eq. 3}
\end{equation}
Since system-environment coupling in the polaron frame is weak, the initial system-environment state could be written as $\rho^{(0)}(0) = \rho_{S}^{(0)}(0)\otimes\rho_{B1}^{(0)}(0)\otimes\rho_{B2}^{(0)}(0)$, where $\rho_{S}^{(0)}(0) = \ket{\uparrow}\bra{\uparrow}$, $\rho_{B1}^{(0)}(0) = \frac{e^{-\beta H_{B1}}}{Z_{B1}}$ with $Z_{B1} = \mathrm{Tr}_{B1}\left(e^{-\beta H_{B1}}\right)$, and $\rho_{B2}^{(0)}(0) = \frac{e^{-\beta H_{B2}}}{Z_{B2}}$ with $Z_{B2} = \mathrm{Tr}_{B2}\left(e^{-\beta H_{B2}}\right)$. Using time-dependent perturbation theory~\cite{koshino2005quantum}, we find that
\newpage
\begin{widetext}
\begin{equation}
\begin{aligned}[b]
\rho_{S}^{(0)}(\tau) = U_{S}^{(0)}(\tau)\bigg[&\rho_{S}^{(0)}(0)+\iota\sum_{\mu}\int_{0}^{\tau}dt_{1}\left[\rho_{S}^{(0)}(0), \widetilde{F}_{\mu}^{(0)}(t_{1})\right]\left\langle\widetilde{B}_{\mu}^{(0)}(t_{1})\right\rangle_{B1}\left\langle\widetilde{J}_{\mu}^{(0)}(t_{1})\right\rangle_{B2}\\
&+\iota\frac{\Delta}{2}\sum_{\mu}\int_{0}^{\tau}dt_{1}\left[\rho_{S}^{(0)}(0), \widetilde{F}_{\mu}^{(0)}(t_{1})\right]\left\langle\widetilde{B}_{\mu}^{(0)}(t_{1})\right\rangle_{B1}\\
&+\sum_{\mu\nu}\int_{0}^{\tau}dt_{1}\int_{0}^{t_{1}}dt_{2}\left(\left[\widetilde{F}_{\mu}^{(0)}(t_{1}), \rho_{S}^{(0)}(0)\widetilde{F}_{\nu}^{(0)}(t_{2})\right]C_{\mu\nu}^{(0)}(t_{1}, t_{2})K_{\mu\nu}^{(0)}(t_{1}, t_{2})+\mathrm{h.c.}\right)\\
&+\frac{\Delta}{2}\sum_{\mu\nu}\int_{0}^{\tau}dt_{1}\int_{0}^{t_{1}}dt_{2}\left(\left[\widetilde{F}_{\mu}^{(0)}(t_{1}), \rho_{S}^{(0)}(0)\widetilde{F}_{\nu}^{(0)}(t_{2})\right]C_{\mu\nu}^{(0)}(t_{1}, t_{2})\left\langle\widetilde{J}_{\nu}^{(0)}(t_{2})\right\rangle_{B2}+\mathrm{h.c.}\right)\\
&+\frac{\Delta}{2}\sum_{\mu\nu}\int_{0}^{\tau}dt_{1}\int_{0}^{t_{1}}dt_{2}\left(\left[\widetilde{F}_{\mu}^{(0)}(t_{1}), \rho_{S}^{(0)}(0)\widetilde{F}_{\nu}^{(0)}(t_{2})\right]C_{\mu\nu}^{(0)}(t_{1}, t_{2})\left\langle\widetilde{J}_{\mu}^{(0)}(t_{1})\right\rangle_{B2}+\mathrm{h.c.}\right)\\
&+\frac{\Delta^{2}}{4}\sum_{\mu\nu}\int_{0}^{\tau}dt_{1}\int_{0}^{t_{1}}dt_{2}\left(\left[\widetilde{F}_{\mu}^{(0)}(t_{1}), \rho_{S}^{(0)}(0)\widetilde{F}_{\nu}^{(0)}(t_{2})\right]C_{\mu\nu}^{(0)}(t_{1}, t_{2})+\mathrm{h.c.}\right)\bigg]U_{S}^{(0)\dagger}(\tau).
\end{aligned}
\label{eq. 4}
\end{equation}
\end{widetext}
Here, $F_{1}^{(0)} = \sigma_{+}$, $F_{2}^{(0)} = \sigma_{-}$, $B_{1}^{(0)} = X$, $B_{2}^{(0)} = X^{\dagger}$, $J_{1}^{(0)} = \sum_{k}\left(f_{k}^{*}c_{k}+f_{k}c_{k}^{\dagger}\right)$, and $J_{2}^{(0)} = \sum_{k}\left(f_{k}^{*}c_{k}+f_{k}c_{k}^{\dagger}\right)$. As to their time-evolved counterparts, $\widetilde{F}_{\mu}^{(0)}(t) = U_{S}^{(0)\dagger}(t)F_{\mu}^{(0)}U_{S}^{(0)}(t)$ with $U_{S}^{(0)}(t) = e^{-\iota H_{S1}t}$, $\widetilde{B}_{\mu}^{(0)}(t) = U_{B1}^{(0)\dagger}(t)B_{\mu}^{(0)}U_{B1}^{(0)}(t)$ with $U_{B1}^{(0)}(t) = e^{-\iota H_{B1}t}$, and $\widetilde{J}_{\mu}^{(0)}(t) = U_{B2}^{(0)\dagger}(t)J_{\mu}^{(0)}U_{B2}^{(0)}(t)$ with $U_{B2}^{(0)}(t) = e^{-\iota H_{B2}t}$. Finally, $\langle\ldots\rangle_{B}$ stands for $\mathrm{Tr}_{B}\left[\rho_{B}^{(0)}(\ldots)\right]$, environment correlation functions are defined as $C_{\mu\nu}^{(0)}(t_{1}, t_{2}) = \left\langle\widetilde{B}_{\mu}^{(0)}(t_{1})\widetilde{B}_{\nu}^{(0)}(t_{2})\right\rangle_{B1}$ and $K_{\mu\nu}^{(0)}(t_{1}, t_{2}) = \left\langle\widetilde{J}_{\mu}^{(0)}(t_{1})\widetilde{J}_{\nu}^{(0)}(t_{2})\right\rangle_{B2}$, and h.c. denotes the Hermitian conjugate.

Now, since we want the modified decay rate, we go one step further and compute the system density matrix obtained with the removal of the evolution effected by the system Hamiltonian. In the polaron frame, we call this matrix $\rho_{Sn}^{(0)}(\tau)$, and it happens to be
\begin{equation*}
\mathrm{Tr}_{B1, B2}\left(e^{\iota H_{S,P}^{(0)}\tau}e^{-\iota H^{(0)}\tau}\rho^{(0)}(0)e^{\iota H^{(0)}\tau}e^{-\iota H_{S,P}^{(0)}\tau}\right),
\end{equation*}
where $H_{S, P}^{(0)} = \frac{\epsilon}{2}\sigma_{z}+\frac{\Delta}{2}(\sigma_{+}e^{\chi}+\sigma_{-}e^{-\chi})$. We note that $e^{\iota H_{S,P}^{(0)}\tau}$ and $e^{-\iota H_{S,P}^{(0)}\tau}$ remove the evolution due to the system Hamiltonian before a measurement is performed. Since we assume that the tunneling amplitude and the system-environment coupling in the polaron frame are small, we could expand both $e^{-\iota H_{S, P}^{(0)}\tau}$ and $e^{-\iota H^{(0)}\tau}$ into perturbation series. Keeping terms to second order then, we find that $\rho_{Sn}^{(0)}(\tau)$ is the sum of $\rho_{S}^{(0)}(\tau)$ and some additional terms. It could easily be shown that most of these additional terms contribute nothing to the modified decay rate. The terms that need to be worked out, however, are
\begin{equation*}
\begin{aligned}[b]
&\mathrm{Tr}_{B1, B2}\left(U^{(0)}(\tau)A_{1}^{(0)}\rho^{(0)}(0)U^{(0)\dagger}(\tau)A_{SP1}^{(0)}\right),\\
&\mathrm{Tr}_{B1, B2}\left(A_{SP1}^{(0)\dagger}U^{(0)}(\tau)\rho^{(0)}(0)A_{1}^{(0)\dagger}U^{(0)\dagger}(\tau)\right),\\
&\mathrm{Tr}_{B1, B2}\left(A_{SP1}^{(0)\dagger}U^{(0)}(\tau)\rho^{(0)}(0)U^{(0)\dagger}(\tau)A_{SP1}^{(0)}\right),\\
&\mathrm{Tr}_{B1, B2}\left(U^{(0)}(\tau)A_{1d}^{(0)}\rho^{(0)}(0)U^{(0)\dagger}(\tau)A_{SP1}^{(0)}\right),\\
\end{aligned}
\end{equation*}
and
\begin{equation*}
\mathrm{Tr}_{B1, B2}\left(A_{SP1}^{(0)\dagger}U^{(0)}(\tau)\rho^{(0)}(0)A_{1d}^{(0)\dagger}U^{(0)\dagger}(\tau)\right)
\end{equation*}
with $U^{(0)}(\tau) = U_{S}^{(0)}(\tau)U_{B1}^{(0)}(\tau)U_{B2}^{(0)}(\tau)$, $A_{1}^{(0)} = -\iota\sum_{\mu}\int_{0}^{\tau}dt\widetilde{F}_{\mu}(t)\otimes\widetilde{B}_{\mu}(t)\otimes\widetilde{J}_{\mu}(t)$, $A_{1d}^{(0)} = -\iota\frac{\Delta}{2}\sum_{\mu}\int_{0}^{\tau}dt\widetilde{F}_{\mu}(t)\otimes\widetilde{B}_{\mu}(t)$, and $A_{SP1}^{(0)} = -\iota\frac{\Delta}{2}\sum_{\mu}\int_{0}^{\tau}dt\widetilde{F}_{\mu}(t)\otimes B_{\mu}$.

\subsubsection{\label{sec. 2ab}Survival probability and modified decay rate}
At time $t = 0$, we prepare our system in the state $\ket{\uparrow}$, where $\sigma_{z}\ket{\uparrow} = \ket{\uparrow}$, and we subsequently perform a measurement after every interval of duration $\tau$ to check if the system state is still $\ket{\uparrow}$. The evolution due to the system Hamiltonian is removed before each measurement is performed so that the survival probability obtained could be used to calculate the modified decay rate. The required survival probability then is $s_{n}^{(0)}(\tau) = 1-\bra{\downarrow}\rho_{Sn}^{(0)}(\tau)\ket{\downarrow}$. Using our expression for $\rho_{S}^{(0)}(\tau)$ (Eq.~(\ref{eq. 4}))---together with the additional terms we computed---to calculate $\rho_{Sn}^{(0)}(\tau)$, we obtain
\begin{widetext}
\begin{equation}
\begin{aligned}[b]
s_{n}^{(0)}(\tau) =\:&1-2\Re\left(\int_{0}^{\tau}dt_{1}\int_{0}^{t_{1}}dt_{2}C_{12}^{(0)}(t_{1}, t_{2})K_{12}^{(0)}(t_{1}, t_{2})e^{\iota\epsilon(t_{1}-t_{2})}\right)\\
&-\frac{\Delta^{2}}{2}\Re\left(\int_{0}^{\tau}dt_{1}\int_{0}^{t_{1}}dt_{2}C_{12}^{(0)}(t_{1}, t_{2})e^{\iota\epsilon(t_{1}-t_{2})}\right)\\
&-\frac{\Delta^{2}}{4}\int_{0}^{\tau}dt_{1}\int_{0}^{\tau}dt_{2}C_{12}^{(0)}(0, 0)e^{\iota\epsilon(t_{2}-t_{1})}\\
&+\frac{\Delta^{2}}{2}\Re\left(\int_{0}^{\tau}dt_{1}\int_{0}^{\tau}dt_{2}C_{12}^{(0)}(\tau, t_{1})e^{\iota\epsilon(t_{2}-t_{1})}\right)\\
=\:&1-2\Re\left(\int_{0}^{\tau}dt_{1}\int_{0}^{t_{1}}dt_{2}C^{(0)}(t_{1}-t_{2})K^{(0)}(t_{1}-t_{2})e^{\iota\epsilon(t_{1}-t_{2})}\right)\\
&-\frac{\Delta^{2}}{2}\Re\left(\int_{0}^{\tau}dt_{1}\int_{0}^{t_{1}}dt_{2}C^{(0)}(t_{1}-t_{2})e^{\iota\epsilon(t_{1}-t_{2})}\right)\\
&-\frac{\Delta^{2}}{4}\int_{0}^{\tau}dt_{1}\int_{0}^{\tau}dt_{2}C^{(0)}(0)e^{\iota\epsilon(t_{2}-t_{1})}\\
&+\frac{\Delta^{2}}{2}\Re\left(\int_{0}^{\tau}dt_{1}\int_{0}^{\tau}dt_{2}C^{(0)}(\tau-t_{1})e^{\iota\epsilon(t_{2}-t_{1})}\right)\\
=\:&1-2\Re\left[\int_{0}^{\tau}dt\int_{0}^{t}dt'\left(C^{(0)}(t')K^{(0)}(t')e^{\iota\epsilon t'}+\frac{\Delta^{2}}{4}C^{(0)}(t')e^{\iota\epsilon t'}\right)\right]\\
&-\frac{\Delta^{2}}{4}\int_{0}^{\tau}dt_{1}\int_{0}^{\tau}dt_{2}C^{(0)}(0)e^{\iota\epsilon(t_{2}-t_{1})}\\
&+\frac{\Delta^{2}}{2}\Re\left(\int_{0}^{\tau}dt\int_{0}^{\tau}dt'C^{(0)}(t')e^{\iota\epsilon(t-\tau+t')}\right).
\end{aligned}
\label{eq. 5}
\end{equation}
\end{widetext}
In the penultimate step, we use $C^{(0)}(t_{1}-t_{2}) = C_{12}^{(0)}(t_{1}, t_{2})$ and $K^{(0)}(t_{1}-t_{2}) = K_{12}^{(0)}(t_{1}, t_{2})$, and in the last step, we change variables via $t' = t_{1}-t_{2}$ and $t = t_{1}$ in the first double integral and via $t' = \tau-t_{1}$ and $t = t_{2}$ in the third one. When we calculate the environment correlation functions, we find that $C^{(0)}(t) = e^{-\Phi_{R1}(t)}e^{-\iota\Phi_{I1}(t)}$ with $\Phi_{R1}(t) = \int_{0}^{\infty}d\omega J(\omega)\frac{4-4\cos(\omega t)}{\omega^{2}}\coth(\frac{\beta\omega}{2})$ and $\Phi_{I1}(t) = \int_{0}^{\infty}d\omega J(\omega)\frac{4\sin(\omega t)}{\omega^2}$ and that $K^{(0)}(t) = \Phi_{R2}(t)-\iota\Phi_{I2}(t)$ with $\Phi_{R2}(t) = \int_{0}^{\infty}d\alpha H(\alpha)\cos(\alpha t)\coth(\frac{\beta\alpha}{2})$ and $\Phi_{I2}(t) = \int_{0}^{\infty}d\alpha H(\alpha)\sin(\alpha t)$. Here, spectral densities have been introduced as $\sum_{k}\abs{g_k}^{2}(\ldots)\,\to\,\int_{0}^{\infty}d\omega J(\omega)(\ldots)$ and $\sum_{k}\abs{f_k}^{2}(\ldots)\,\to\,\int_{0}^{\infty}d\alpha H(\alpha)(\ldots)$.

Since system-environment coupling in the polaron frame is weak, we could neglect the build-up of correlations between the system and the environment and write the survival probability at time $t = N\tau$ as $S_{n}^{(0)}(t = N\tau) = \left(s_{n}^{(0)}(\tau)\right)^{N} \equiv e^{-\Gamma^{(0)}_{n}(\tau)N\tau}$, thereby defining the modified decay rate $\Gamma_{n}^{(0)}(\tau)$. It follows that $\Gamma_{n}^{(0)}(\tau) = -\frac{1}{\tau}\ln s_{n}^{(0)}(\tau)$~\cite{chaudhry2017quantum}. Using the expression we derived for $s_{n}^{(0)}(\tau)$ (Eq.~(\ref{eq. 5})), we can work out the following expression for $\Gamma_{n}^{(0)}(\tau)$:
\begin{equation*}
\begin{aligned}[b]
&\frac{2}{\tau}\int_{0}^{\tau}dt\int_{0}^{t}dt'e^{-\Phi_{R1}\left(t'\right)}\cos(\epsilon t'-\Phi_{I1}(t'))\Phi_{R2}(t')\\
&+\frac{2}{\tau}\int_{0}^{\tau}dt\int_{0}^{t}dt'e^{-\Phi_{R1}\left(t'\right)}\sin(\epsilon t'-\Phi_{I1}(t'))\Phi_{I2}(t')\\
&+\frac{\Delta^{2}}{2\tau}\int_{0}^{\tau}dt\int_{0}^{t}dt'e^{-\Phi_{R1}\left(t'\right)}\cos(\epsilon t'-\Phi_{I1}(t'))\\
&+\frac{\Delta^{2}}{\tau\epsilon^{2}}\sin^{2}\left(\frac{\epsilon\tau}{2}\right)e^{-\Phi_{R1}(0)-\iota\Phi_{I1}(0)}\\
&-\frac{\Delta^{2}}{\tau\epsilon}\sin\left(\frac{\epsilon\tau}{2}\right)\int_{0}^{\tau}dte^{-\Phi_{R1}(t)}\cos\left[\epsilon\left(t-\frac{\tau}{2}\right)-\Phi_{I1}(t)\right].
\end{aligned}
\end{equation*}

We now plot $\Gamma_{n}^{(0)}(\tau)$ against $\tau$. To do so, we model the spectral densities as $J(\omega) = G\omega^{s}\omega_{c}^{1-s}e^{-\omega/\omega_{c}}$ and $H(\alpha) = F\alpha^{r}\alpha_{c}^{1-r}e^{-\alpha/\alpha_{c}}$, where $G$ and $F$ are dimensionless parameters characterizing the system-environment coupling strengths, $\omega_{c}$ and $\alpha_{c}$ are cut-off frequencies, and $s$ and $r$ are Ohmicity parameters~\cite{breuer2002theory}. Whereas $G$ corresponds to strong coupling, $F$ corresponds to the weak one; therefore, in our plots, we keep $G$ greater than $F$. To be particular, we work at zero temperature and look at the Ohmic case for each of the spectral densities ($s = 1$ and $r = 1$). Doing so gives $\Phi_{R1}(t) = 2G\ln(1+\omega_{c}^{2}t^{2})$, $\Phi_{I1}(t) = 4G\tan^{-1}(\omega_{c}t)$, $\Phi_{R2}(t) = F\frac{\alpha_{c}^2\left(1-\alpha_{c}^{2}t^{2}\right)}{\left(1+\alpha_{c}^{2}t^{2}\right)^{2}}$, and $\Phi_{I2}(t) = 2F\frac{\alpha_{c}^{3}t}{\left(1+\alpha_{c}^{2}t^{2}\right)^{2}}$, allowing us to write our expression for $\Gamma_{n}^{(0)}(\tau)$ as
\begin{equation*}
\begin{aligned}[b]
&\frac{2F}{\tau}\int_{0}^{\tau}dt\int_{0}^{t}dt'\frac{\alpha_{c}^{2}\left(1-\alpha_{c}^{2}t'^{2}\right)\cos(\epsilon t'-4G\tan^{-1}(\omega_{c}t'))}{\left(1+\alpha_{c}^{2}t'^{2}\right)^{2}\left(1+\omega_{c}^{2}t'^2\right)^{2G}}\\
&+\frac{4F}{\tau}\int_{0}^{\tau}dt\int_{0}^{t}dt'\frac{\alpha_{c}^{3}t'\sin(\epsilon t'-4G\tan^{-1}(\omega_{c}t'))}{\left(1+\alpha_{c}^{2}t'^{2}\right)^{2}\left(1+\omega_{c}^{2}t'^2\right)^{2G}}\\
&+\frac{2}{\tau}\int_{0}^{\tau}dt\int_{0}^{t}dt'\frac{\Delta^{2}}{4}\frac{\cos(\epsilon t'-4G\tan^{-1}(\omega_{c}t'))}{\left(1+\omega_{c}^{2}t'^2\right)^{2G}}\\
&+\frac{\Delta^{2}}{\tau\epsilon^{2}}\sin^{2}\left(\frac{\epsilon\tau}{2}\right)\\
&-\frac{\Delta^{2}}{\tau\epsilon}\sin\left(\frac{\epsilon\tau}{2}\right)\int_{0}^{\tau}dt\frac{\cos[\epsilon(t-\tau/2)-4G\tan^{-1}(\omega_{c}t)]}{(1+\omega_{c}^{2}t^{2})}.
\end{aligned}
\end{equation*}

\begin{figure}[H]
\centering
\begin{subfigure}[H]{0.48\textwidth}
\centering
\includegraphics[width = \textwidth]{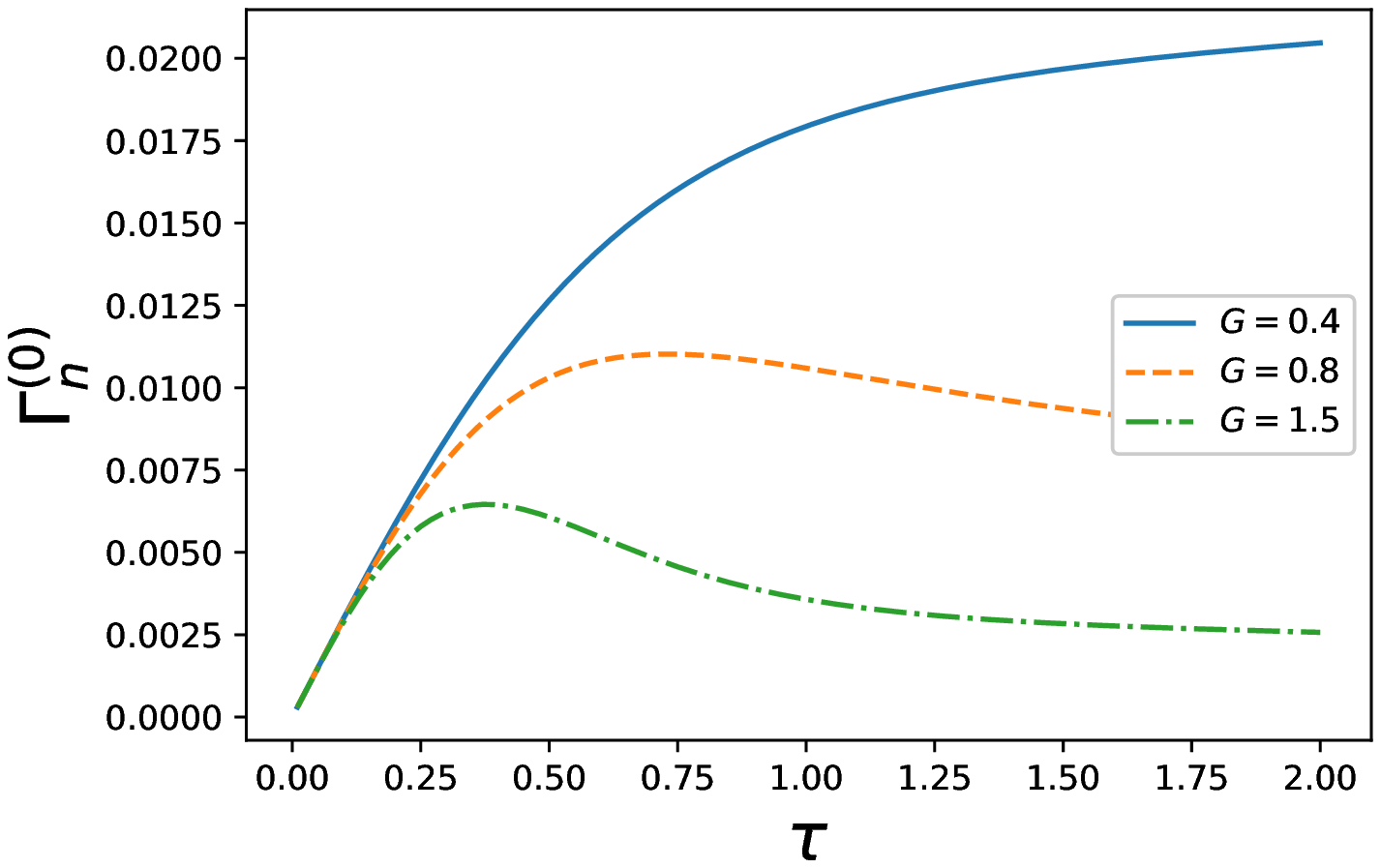}
\caption{}
\label{fig. 1a}
\end{subfigure}
\hfill
\begin{subfigure}[H]{0.48\textwidth}
\centering
\includegraphics[width = \textwidth]{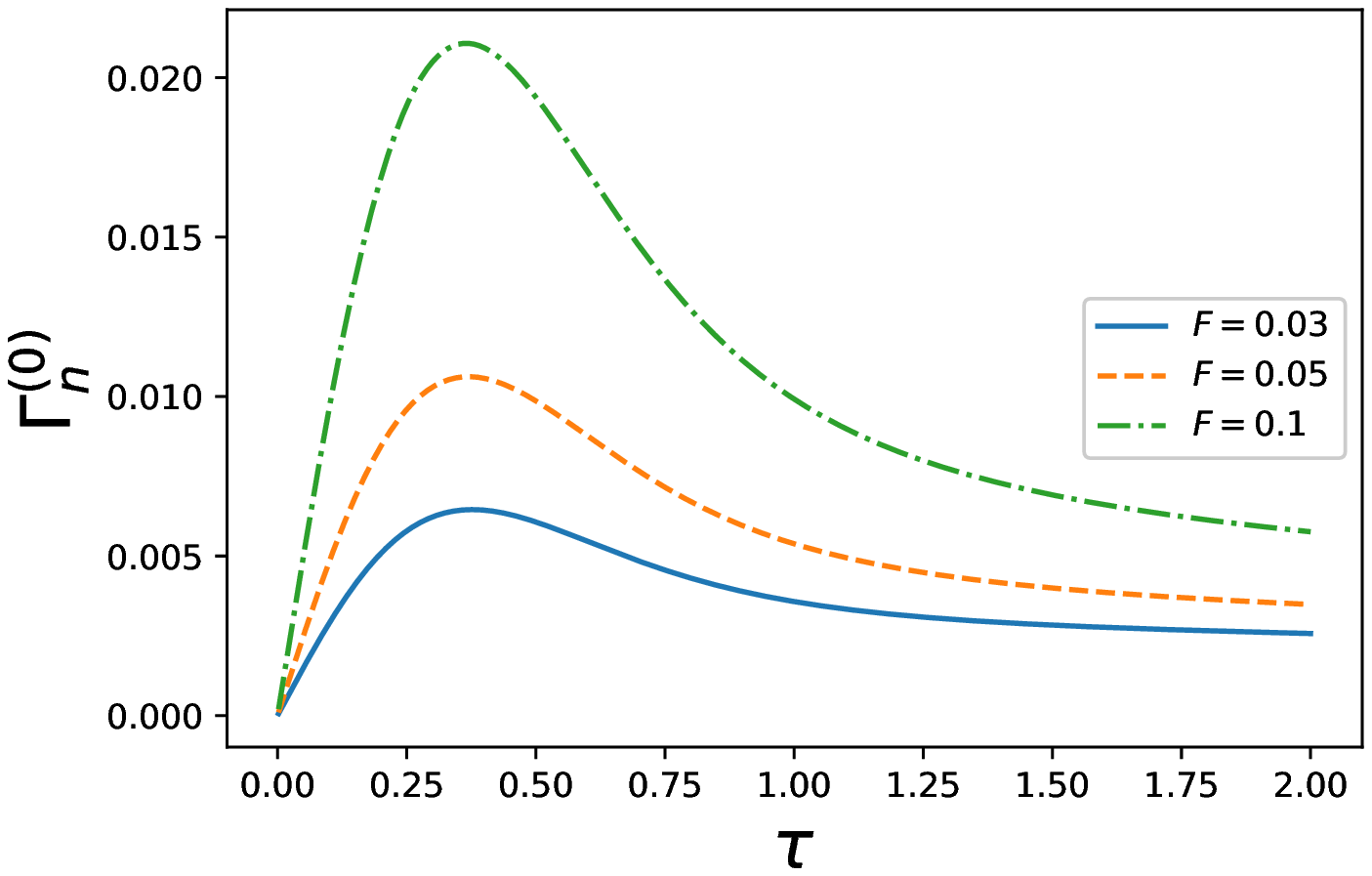}
\caption{}
\label{fig. 1b}
\end{subfigure}
\caption{\textbf{Variation of the modified decay rate for the spin-boson model with both strong and weak system-environment coupling strengths.} (a) Graph of $\Gamma^{(0)}$ (at zero temperature) against $\tau$ when $G = 0.4$ (solid, blue curve), $G = 0.8$ (dashed, orange curve), and $G = 1.5$ (dot-dashed, green curve). Here, we have used an Ohmic environment ($s = 1$ and $r = 1$) with $F = 0.03$, $\epsilon = 1$, $\omega_{c} = 1$, $\alpha_{c} = 1$, and $\Delta = 0.05$. The initial state is $\ket{\uparrow}$. (b) Graph of $\Gamma^{(0)}$ (at zero temperature) against $\tau$ when $F = 0.03$ (solid, blue curve), $F = 0.05$ (dashed, orange curve), and $F = 0.1$ (dot-dashed, green curve). Here, we have again used an Ohmic environment ($s = 1$ and $r = 1$) with $G = 1.5$, $\epsilon = 1$, $\omega_{c} = 1$, $\alpha_{c} = 1$, and $\Delta = 0.05$. The initial state is still $\ket{\uparrow}$. $\hbar$ is equal to $1$ throughout.}
\label{fig. 1}
\end{figure}

The integrals above could be worked out numerically, and results (graphs of $\Gamma_{n}^{(0)}(\tau)$ against $\tau$) are shown in Fig.~\ref{fig. 1} for different values of the system-environment coupling strengths, $G$ and $F$. What is absolutely clear is that despite having removed the system Hamiltonian evolution, we obtain the results of Ref.~\cite{chaudhry2017quantum}: increasing the strong coupling strength leads to a decrease in the decay rate (Fig.~\ref{fig. 1a}) whereas increasing the weak coupling strength leads to an increase (Fig.~\ref{fig. 1b}). Also, just as Ref.~\cite{chaudhry2017quantum} illustrates, whereas increasing the weak coupling strength does not change the qualitative behavior of the Zeno/anti-Zeno transition, increasing the strong coupling strength does have an effect, namely that it causes the transition to occur at smaller values of $\tau$. Clearly, system evolution has no bearing on the general effects of strong and weak system-environment couplings.
\null
\vfill
\vfill
\newpage

\subsection{\label{sec. 2b}Large spin-boson model}
\subsubsection{\label{sec. 2ba}System density matrix}
The Hamiltonian to be considered for this case is $H_{L}^{(1)}$ (Eq.~(\ref{eq. 2})), and we again use a polaron transformation to transform it, the transformation being the same as that used in section~\ref{sec. 2aa}. The Hamiltonian in the polaron frame is thus
\begin{equation}
\begin{aligned}[b]
H^{(1)} =\:&\epsilon J_{z}+\sum_{k}\omega_{k}b_{k}^{\dagger}b_{k}+\sum_{k}\alpha_{k}c_{k}^{\dagger}c_{k}-\kappa J_{z}^{2}\\
&+\left[\frac{\Delta}{2}+\sum_{k}\left(f_{k}^{*}c_{k}+f_{k}c_{k}^{\dagger}\right)\right]\left(J_{+}e^{\chi}+J_{-}e^{-\chi}\right),
\end{aligned}
\label{eq. 6}
\end{equation}
where $\kappa = 4\sum_{k}\frac{\abs{g_{k}}^{2}}{\omega_{k}}$.

Since the system-environment coupling in the polaron frame is weak again, we could write the initial system-environment state as $\rho^{(1)}(0)$ = $\rho_{S}^{(1)}(0)\otimes\rho_{B1}^{(1)}(0)\otimes\rho_{B2}^{(1)}(0)$, where $\rho_{S}^{(1)}(0) = \ket{j}\bra{j}$, $\rho_{B1}^{(1)}(0) = \frac{e^{-\beta H'_{B1}}}{Z'_{B1}}$ with $Z'_{B1} = \mathrm{Tr}_{B1}\left(e^{-\beta H'_{B1}}\right)$, and $\rho_{B2}^{(1)}(0) = \frac{e^{-\beta H'_{B2}}}{Z'_{B2}}$ with $Z'_{B2} = \mathrm{Tr}_{B2}\left(e^{-\beta H'_{B2}}\right)$. $\ket{j}$ represents a spin coherent state with $j = N_{S}/2$, where $N_{S}$, as said in section~\ref{sec. 1}, is the number of two-level systems we work with in our large spin-boson model. Using time-dependent perturbation theory then, we find that
\begin{widetext}
\begin{equation}
\begin{aligned}[b]
\rho_{S}^{(1)}(\tau) = U_{S}^{(1)}(\tau)\bigg[&\rho_{S}^{(1)}(0)+\iota\sum_{\mu}\int_{0}^{\tau}dt_{1}\left[\rho_{S}^{(1)}(0), \widetilde{F}_{\mu}^{(1)}(t_{1})\right]\left\langle\widetilde{B}_{\mu}^{(1)}(t_{1})\right\rangle_{B1}\left\langle\widetilde{J}_{\mu}^{(1)}(t_{1})\right\rangle_{B2}\\
&+\iota\frac{\Delta}{2}\sum_{\mu}\int_{0}^{\tau}dt_{1}\left[\rho_{S}^{(1)}(0), \widetilde{F}_{\mu}^{(1)}(t_{1})\right]\left\langle\widetilde{B}_{\mu}^{(1)}(t_{1})\right\rangle_{B1}\\
&+\sum_{\mu\nu}\int_{0}^{\tau}dt_{1}\int_{0}^{t_{1}}dt_{2}\left(\left[\widetilde{F}_{\mu}^{(1)}(t_{1}), \rho_{S}^{(1)}(0)\widetilde{F}_{\nu}^{(1)}(t_{2})\right]C_{\mu\nu}^{(1)}(t_{1}, t_{2})K_{\mu\nu}^{(1)}(t_{1}, t_{2})+\mathrm{h.c.}\right)\\
&+\frac{\Delta}{2}\sum_{\mu\nu}\int_{0}^{\tau}dt_{1}\int_{0}^{t_{1}}dt_{2}\left(\left[\widetilde{F}_{\mu}^{(1)}(t_{1}), \rho_{S}^{(1)}(0)\widetilde{F}_{\nu}^{(1)}(t_{2})\right]C_{\mu\nu}^{(1)}(t_{1}, t_{2})\left\langle\widetilde{J}_{\nu}^{(1)}(t_{2})\right\rangle_{B2}+\mathrm{h.c.}\right)\\
&+\frac{\Delta}{2}\sum_{\mu\nu}\int_{0}^{\tau}dt_{1}\int_{0}^{t_{1}}dt_{2}\left(\left[\widetilde{F}_{\mu}^{(1)}(t_{1}), \rho_{S}^{(1)}(0)\widetilde{F}_{\nu}^{(1)}(t_{2})\right]C_{\mu\nu}^{(1)}(t_{1}, t_{2})\left\langle\widetilde{J}_{\mu}^{(1)}(t_{1})\right\rangle_{B2}+\mathrm{h.c.}\right)\\
&+\frac{\Delta^{2}}{4}\sum_{\mu\nu}\int_{0}^{\tau}dt_{1}\int_{0}^{t_{1}}dt_{2}\left(\left[\widetilde{F}_{\mu}^{(1)}(t_{1}), \rho_{S}^{(1)}(0)\widetilde{F}_{\nu}^{(1)}(t_{2})\right]C_{\mu\nu}^{(1)}(t_{1}, t_{2})+\mathrm{h.c.}\right)\bigg]U_{S}^{(1)\dagger}(\tau).
\end{aligned}
\label{eq. 7}
\end{equation}
\end{widetext}
Here, $F_{1}^{(1)} = J_{+}$, $F_{2}^{(1)} = J_{-}$, $B_{1}^{(1)} = X$, $B_{2}^{(1)} = X^{\dagger}$, $J_{1}^{(1)} = \sum_{k}\left(f_{k}^{*}c_{k}+f_{k}c_{k}^{\dagger}\right)$, and $J_{2}^{(1)} = \sum_{k}\left(f_{k}^{*}c_{k}+f_{k}c_{k}^{\dagger}\right)$. Their time-evolved counterparts happen to be $\widetilde{F}_{\mu}^{(1)}(t) = U_{S}^{(1)\dagger}(t)F_{\mu}^{(1)}U_{S}^{(1)}(t)$ with $U_{S}^{(1)}(t) = e^{-\iota\left(H'_{S1}-\kappa J_{z}^{2}\right)t}$, $\widetilde{B}_{\mu}^{(1)}(t) = U_{B1}^{(1)\dagger}(t)B_{\mu}^{(1)}U_{B1}^{(1)}(t)$ with $U_{B1}^{(1)}(t) = e^{-\iota H'_{B1}t}$, and $\widetilde{J}_{\mu}^{(1)}(t) = U_{B2}^{(1)\dagger}(t)J_{\mu}^{(1)}U_{B2}^{(1)}(t)$ with $U_{B2}^{(1)}(t) = e^{-\iota H'_{B2}t}$. Finally, $\langle\ldots\rangle_{B}$ stands for $\mathrm{Tr}_{B}\left[\rho_{B}^{(1)}(\ldots)\right]$, environment correlation functions are defined as $C_{\mu\nu}^{(1)}(t_{1}, t_{2}) = \left\langle\widetilde{B}_{\mu}^{(1)}(t_{1})\widetilde{B}_{\nu}^{(1)}(t_{2})\right\rangle_{B1}$ and $K_{\mu\nu}^{(1)}(t_{1}, t_{2}) = \left\langle\widetilde{J}_{\mu}^{(1)}(t_{1})\widetilde{J}_{\nu}^{(1)}(t_{2})\right\rangle_{B2}$, and h.c. denotes the Hermitian conjugate.

Since we want the modified decay rate, however, we compute the system density matrix with the system Hamiltonian evolution removed. Calling this matrix $\rho_{Sn}^{(1)}(\tau)$ in the polaron frame, we find it to be
\begin{equation*}
\mathrm{Tr}_{B1, B2}\left(e^{\iota H_{S,P}^{(1)}\tau}e^{-\iota H^{(1)}\tau}\rho^{(1)}(0)e^{\iota H^{(1)}\tau}e^{-\iota H_{S,P}^{(1)}\tau}\right),
\end{equation*}
where $H_{S, P}^{(1)} = \epsilon J_{z}+\frac{\Delta}{2}(J_{+}X+J_{-}X^{\dagger})$. It could easily be noted that $e^{\iota H_{S, P}^{(1)}\tau}$ and $e^{-\iota H_{S, P}^{(1)}\tau}$ remove the evolution due to the system Hamiltonian before a measurement is performed. Assuming that the tunneling amplitude and the system-environment coupling in the polaron frame are small, we expand both $e^{-\iota H_{S, P}^{(1)}\tau}$ and $e^{-\iota H^{(1)}\tau}$ into perturbation series, and keeping terms to second order only, we find that $\rho_{Sn}^{(1)}(\tau)$ is the sum of $\rho_{S}^{(1)}(\tau)$ and some additional terms. As before, we could show that most of these additional terms do not contribute anything to the modified decay rate. The terms that need to be worked out, however, are
\begin{equation*}
\begin{aligned}[b]
&\mathrm{Tr}_{B1, B2}\left(A_{SP1}^{(1)\dagger}U^{(1)}(\tau)A_{1}^{(1)}\rho^{(1)}(0)U^{(1)\dagger}(\tau)\right),\\
&\mathrm{Tr}_{B1, B2}\left(A_{SP1}^{(1)\dagger}U^{(1)}(\tau)A_{1d}^{(1)}\rho^{(1)}(0)U^{(1)\dagger}(\tau)\right),\\
&\mathrm{Tr}_{B1, B2}\left(A_{SP2}^{(1)\dagger}U^{(1)}(\tau)\rho^{(1)}(0)U^{(1)\dagger}(\tau)\right),\\
&\mathrm{Tr}_{B1, B2}\left(U^{(1)}(\tau)\rho^{(1)}(0)U^{(1)\dagger}(\tau)A_{SP2}^{(1)}\right),\\
&\mathrm{Tr}_{B1, B2}\left(U^{(1)}(\tau)\rho^{(1)}(0)A_{1}^{(1)\dagger}U^{(1)\dagger}(\tau)A_{SP1}^{(1)}\right),\\
&\mathrm{Tr}_{B1, B2}\left(U^{(1)}(\tau)\rho^{(1)}(0)A_{1d}^{(1)\dagger}U^{(1)\dagger}(\tau)A_{SP1}^{(1)}\right),\\
\end{aligned}
\end{equation*}
and
\begin{equation*}
\mathrm{Tr}_{B1, B2}\left(U^{(1)}(\tau)\rho^{(1)}(0)U^{(1)\dagger}(\tau)A_{SP2}^{(1)\dagger}\right)
\end{equation*}
with $U^{(1)}(\tau) = U_{S}^{(1)}(\tau)U_{B1}^{(1)}(\tau)U_{B2}^{(1)}(\tau)$, $A_{1}^{(1)} = -\iota\sum_{\mu}\int_{0}^{\tau}dt\widetilde{F}_{\mu}(t)\otimes\widetilde{B}_{\mu}(t)\otimes\widetilde{J}_{\mu}(t)$, $A_{1d}^{(1)} = -\iota\frac{\Delta}{2}\sum_{\mu}\int_{0}^{\tau}dt\widetilde{F}_{\mu}(t)\otimes\widetilde{B}_{\mu}(t)$, $A_{SP1}^{(1)} = -\iota\frac{\Delta}{2}\sum_{\mu}\int_{0}^{\tau}dt\widetilde{F}_{\mu}(t)\otimes B_{\mu}$, and $A_{SP2}^{(1)} = -\frac{\Delta^{2}}{4}\sum_{\mu\nu}\int_{0}^{\tau}dt_{1}\int_{0}^{\tau}dt_{2}\widetilde{F}_{\mu}(t_{1})\widetilde{F}_{\nu}(t_{2})\otimes B_{\mu}B_{\nu}$.

\subsubsection{\label{sec. 2bb}Survival probability and modified decay rate}
We prepare our system in the state $\ket{j}$, where $J_{z}\ket{j} = j\ket{j}$, at time $t = 0$ and perform a measurement after every interval of duration $\tau$ to check if the system state is still $\ket{j}$. The system Hamiltonian evolution is removed before every measurement so that the survival probability obtained corresponds to the modified decay rate. The required survival probability is thus $s_{n}^{(1)}(\tau) = \bra{j}\rho_{Sn}^{(1)}(\tau)\ket{j}$, where $\rho_{Sn}^{(1)}$ is just the sum of $\rho_{S}^{(1)}(\tau)$ (Eq.~(\ref{eq. 7})) and the additional terms we computed. We hence get
\begin{widetext}
\begin{equation}
\begin{aligned}[b]
s_{n}^{(1)}(\tau) =\:&1-4j\Re\left(\int_{0}^{\tau}dt_{1}\int_{0}^{t_{1}}dt_{2}C_{12}^{(1)}(t_{1}, t_{2})K_{12}^{(1)}(t_{1}, t_{2})e^{\iota\left[\epsilon(t_{1}-t_{2})+\kappa(1-2j)(t_{1}-t_{2})\right]}\right)\\
&-\Delta^{2}j\Re\left(\int_{0}^{\tau}dt_{1}\int_{0}^{t_{1}}dt_{2}C_{12}^{(1)}(t_{1}, t_{2})e^{\iota\left[\epsilon(t_{1}-t_{2})+\kappa(1-2j)(t_{1}-t_{2})\right]}\right)\\
&-\Delta^{2}j\int_{0}^{\tau}dt_{1}\int_{0}^{\tau}dt_{2}C_{12}^{(1)}(0, 0)e^{\iota\epsilon(t_{2}-t_{1})}\\
&+\Delta^{2}j\Re\left(\int_{0}^{\tau}dt_{1}\int_{0}^{\tau}dt_{2}C_{12}^{(1)}(\tau, t_{1})e^{\iota\left[\epsilon(t_{2}-t_{1})+\kappa(2j-1)(-\tau+t_{1})\right]}\right)\\
=\:&1-4j\Re\left(\int_{0}^{\tau}dt_{1}\int_{0}^{t_{1}}dt_{2}C^{(1)}(t_{1}-t_{2})K^{(1)}(t_{1}-t_{2})e^{\iota\left[\epsilon(t_{1}-t_{2})+\kappa(1-2j)(t_{1}-t_{2})\right]}\right)\\
&-\Delta^{2}j\Re\left(\int_{0}^{\tau}dt_{1}\int_{0}^{t_{1}}dt_{2}C^{(1)}(t_{1}-t_{2})e^{\iota\left[\epsilon(t_{1}-t_{2})+\kappa(1-2j)(t_{1}-t_{2})\right]}\right)\\
&-\Delta^{2}j\int_{0}^{\tau}dt_{1}\int_{0}^{\tau}dt_{2}C^{(1)}(0)e^{\iota\epsilon(t_{2}-t_{1})}\\
&+\Delta^{2}j\Re\left(\int_{0}^{\tau}dt_{1}\int_{0}^{\tau}dt_{2}C^{(1)}(\tau-t_{1})e^{\iota\left[\epsilon(t_{2}-t_{1})+\kappa(2j-1)(-\tau+t_{1})\right]}\right)\\
=\:&1-4j\Re\left(\int_{0}^{\tau}dt\int_{0}^{t}dt'C^{(1)}(t')K^{(1)}(t')e^{\iota\left[\epsilon t'+\kappa(1-2j)t'\right]}\right)\\
&-\Delta^{2}j\Re\left(\int_{0}^{\tau}dt\int_{0}^{t}dt'C^{(1)}(t')e^{\iota\left[\epsilon t'+\kappa(1-2j)t'\right]}\right)\\
&-\Delta^{2}j\int_{0}^{\tau}dt_{1}\int_{0}^{\tau}dt_{2}C^{(1)}(0)e^{\iota\epsilon(t_{2}-t_{1})}\\
&+\Delta^{2}j\Re\left(\int_{0}^{\tau}dt\int_{0}^{\tau}dt'C^{(1)}(t')e^{\iota\left[\epsilon(t-\tau+t')-\kappa(2j-1)t'\right]}\right).
\end{aligned}
\label{eq. 8}
\end{equation}
\end{widetext}
In the penultimate step, we use $C^{(1)}(t_{1}-t_{2}) = C_{12}^{(1)}(t_{1}, t_{2})$ and $K^{(1)}(t_{1}-t_{2}) = K_{12}^{(1)}(t_{1}, t_{2})$, and in the last step, we change variables via $t' = t_{1}-t_{2}$ and $t = t_{1}$ in the first two integrals and via $t' = \tau-t_{1}$ and $t = t_{2}$ in the fourth one. Calculating the correlation functions $C^{(1)}(t)$ and $K^{(1)}(t)$, we find that $C^{(1)}(t) = e^{-\Phi_{R1}(t)}e^{-\iota\Phi_{I1}(t)}$ with $\Phi_{R1}(t) = \int_{0}^{\infty}d\omega J(\omega)\frac{4-4\cos(\omega t)}{\omega^{2}}\coth(\frac{\beta\omega}{2})$ and $\Phi_{I1}(t) = \int_{0}^{\infty}d\omega J(\omega)\frac{4\sin(\omega t)}{\omega^2}$ and that $K^{(1)}(t) = \Phi_{R2}(t)-\iota\Phi_{I2}(t)$ with $\Phi_{R2}(t) = \int_{0}^{\infty}d\alpha H(\alpha)\cos(\alpha t)\coth(\frac{\beta\alpha}{2})$ and $\Phi_{I2}(t) = \int_{0}^{\infty}d\alpha H(\alpha)\sin(\alpha t)$, where the spectral density has been introduced as $\sum_{k}\abs{g_k}^{2}(\ldots)\,\to\,\int_{0}^{\infty}d\omega J(\omega)(\ldots)$ and $\sum_{k}\abs{f_k}^{2}(\ldots)\,\to\,\int_{0}^{\infty}d\alpha H(\alpha)(\ldots)$.

Again, since system-environment coupling in the polaron frame is weak, we could ignore the correlations building between the system and the environment and follow the reasoning in section~\ref{sec. 2ab} to show that the modified decay rate, $\Gamma_{n}^{(1)}(\tau)$, is
\begin{equation*}
\begin{aligned}[b]
&\frac{4j}{\tau}\int_{0}^{\tau}dt\int_{0}^{t}dt'e^{-\Phi_{R1}(t')}\cos(D_{1}(t'))\Phi_{R2}(t')\\
&+\frac{4j}{\tau}\int_{0}^{\tau}dt\int_{0}^{t}dt'e^{-\Phi_{R1}(t')}\sin(D_{1}(t'))\Phi_{I2}(t')\\
&+\frac{\Delta^{2}j}{\tau}\int_{0}^{\tau}dt\int_{0}^{t}dt'e^{-\Phi_{R1}(t')}\cos(D_{1}(t'))\\
&+\frac{\Delta^{2}}{\tau}(2j)\frac{1}{\epsilon^{2}}\sin^{2}\left(\frac{\epsilon\tau}{2}\right)e^{-\Phi_{R1}(0)-\iota\Phi_{I1}(0)}\\
&-\frac{\Delta^{2}}{\tau}(2j)\frac{1}{\epsilon}\sin\left(\frac{\epsilon\tau}{2}\right)\int_{0}^{\tau}dte^{-\Phi_{R1}(t)}\cos(D_{2}(t)),
\end{aligned}
\end{equation*}
where $D_{1}(t) = \epsilon t+\kappa(1-2j)t-\Phi_{I1}(t)$ and $D_{2}(t) = -\kappa(2j-1)t+\epsilon(t-\tau/2)-\Phi_{I}(t)$.

To plot $\Gamma_{n}^{(1)}(\tau)$ against $\tau$, we model the spectral densities the same way as before and look at the Ohmic case for each of them. Working at zero temperature then allows us to write our expression for $\Gamma_{n}^{(1)}(\tau)$ as
\begin{equation*}
\begin{aligned}[b]
&\frac{4Fj}{\tau}\int_{0}^{\tau}dt\int_{0}^{t}dt'\frac{\alpha_{c}^{2}(1-\alpha_{c}^{2}t'^{2})\cos(D_{1}(t'))}{(1+\alpha_{c}^{2}t'^{2})^{2}(1+\omega_{c}^{2}t'^{2})^{2G}}\\
&+\frac{8Fj}{\tau}\int_{0}^{\tau}dt\int_{0}^{t}dt'\frac{\alpha_{c}^{3}t'\sin(D_{1}(t'))}{(1+\alpha_{c}^{2}t'^{2})^{2}(1+\omega_{c}^{2}t'^{2})^{2G}}\\
&+\frac{\Delta^{2}j}{\tau}\int_{0}^{\tau}dt\int_{0}^{t}dt'\frac{\cos(D_{1}(t'))}{(1+\omega_{c}^{2}t'^{2})^{2G}}\\
&+\frac{\Delta^{2}}{\tau}(2j)\frac{1}{\epsilon^{2}}\sin^{2}\left(\frac{\epsilon\tau}{2}\right)\\
&-\frac{\Delta^{2}}{\tau}(2j)\frac{1}{\epsilon}\sin\left(\frac{\epsilon\tau}{2}\right)\int_{0}^{\tau}dt\frac{\cos(D_{2}(t))}{(1+\omega_{c}^{2}t^{2})^{2G}}.
\end{aligned}
\end{equation*}

\begin{figure}[H]
\centering
\begin{subfigure}[H]{0.48\textwidth}
\centering
\includegraphics[width = \textwidth]{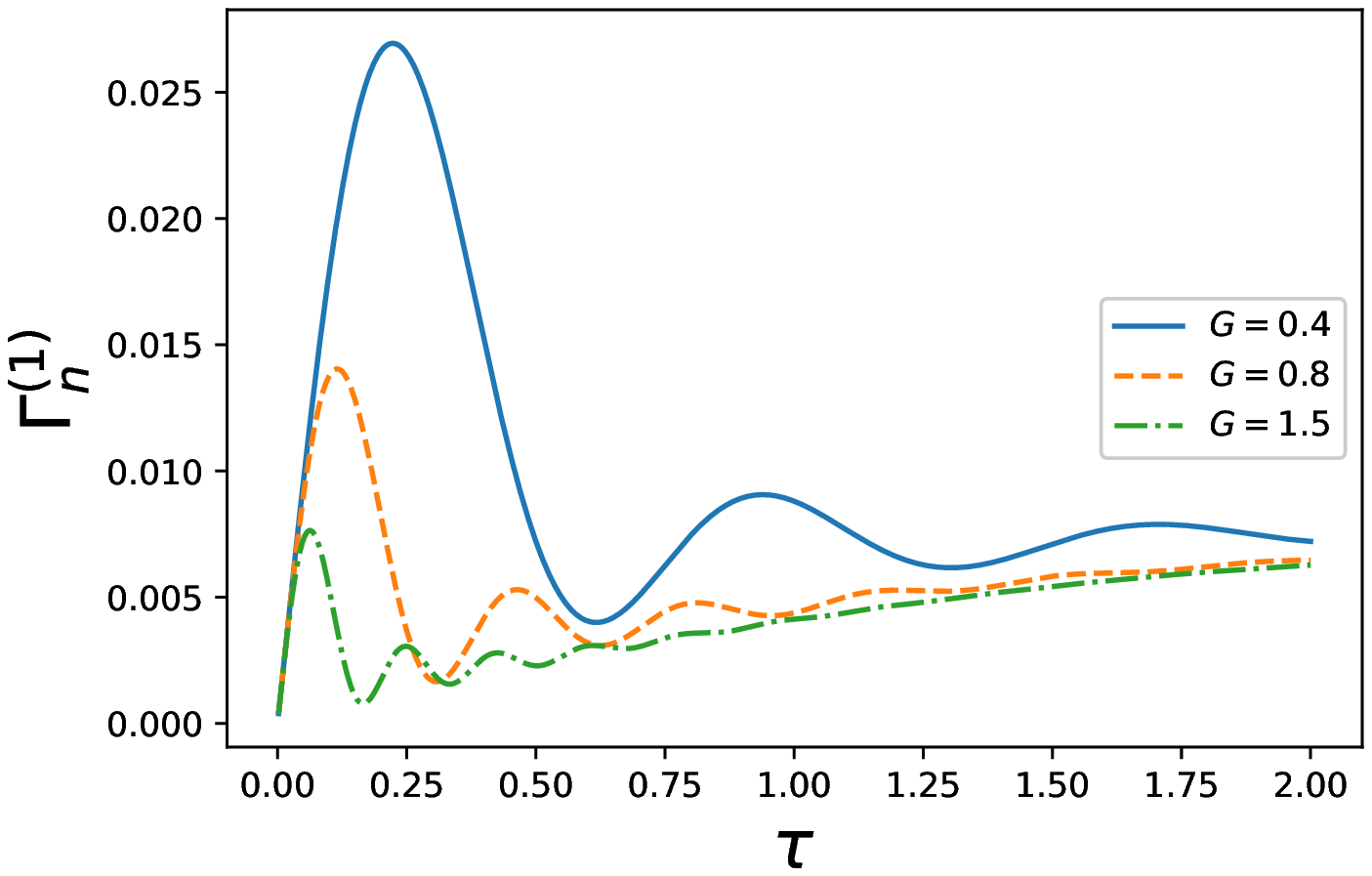}
\caption{}
\label{fig. 2a}
\end{subfigure}
\hfill
\begin{subfigure}[H]{0.48\textwidth}
\centering
\includegraphics[width = \textwidth]{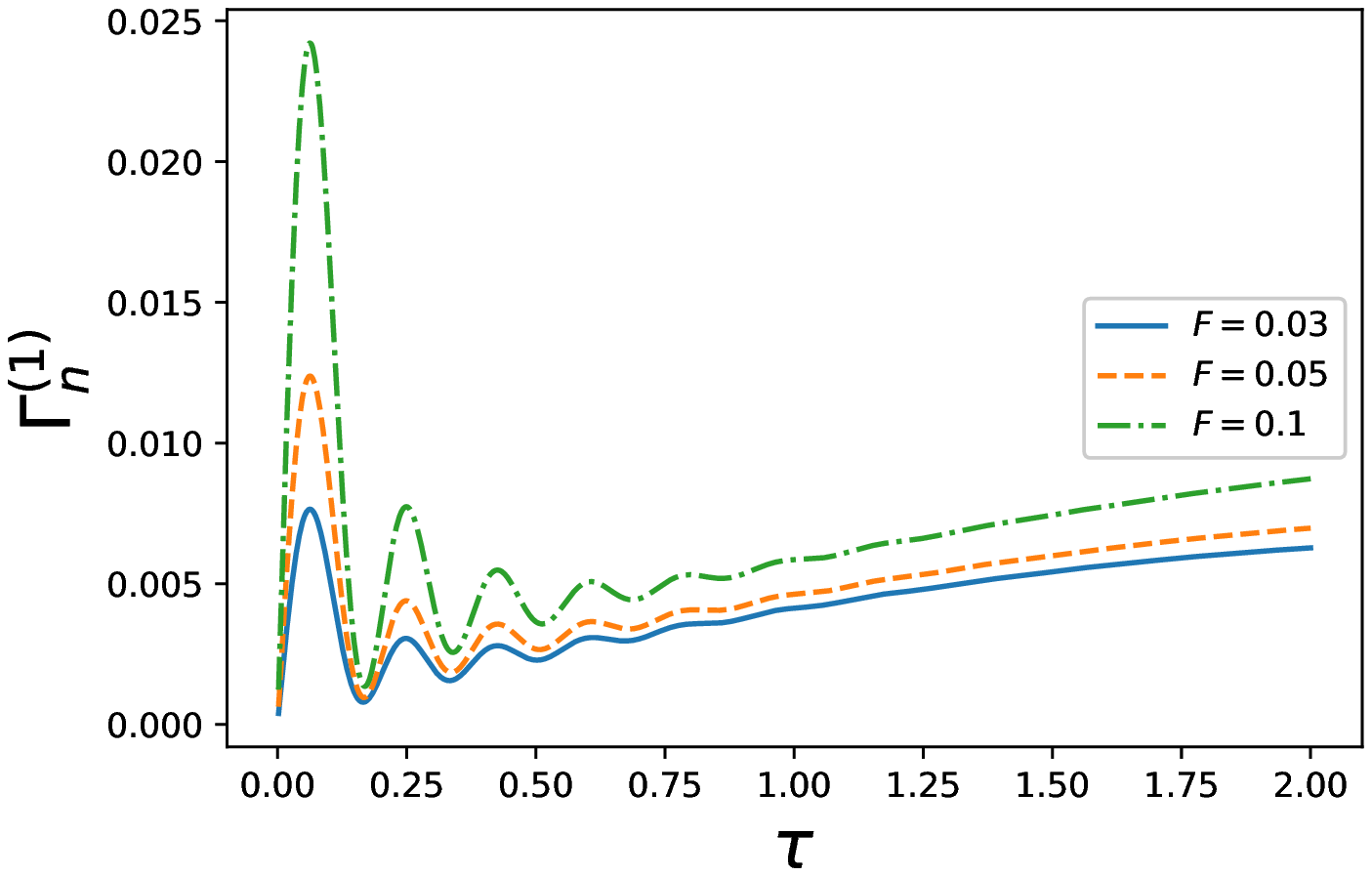}
\caption{}
\label{fig. 2b}
\end{subfigure}
\caption{\textbf{Variation of the modified decay rate for the large spin-boson model with both strong and weak system-environment coupling strengths.} (a) Graph of $\Gamma^{(1)}$ (at zero temperature) against $\tau$ when $G = 0.4$ (solid, blue curve), $G = 0.8$ (dashed, orange curve), and $G = 1.5$ (dot-dashed, green curve). Here, we have used an Ohmic environment ($r = 1$) with $\epsilon = 1$, $\alpha_{c} = 1$, $\Delta = 0.05$, and $j = 3$. The initial state is $\ket{j}$. (b) Graph of $\Gamma^{(1)}$ (at zero temperature) against $\tau$ when $F = 0.03$ (solid, blue curve), $F = 0.05$ (dashed, orange curve), and $F = 0.1$ (dot-dashed, green curve). Here, we have again used an Ohmic environment ($s = 1$ and $r = 1$) with $G = 1.5$, $\epsilon = 1$, $\omega_{c} = 1$, $\alpha_{c} = 1$, $\Delta = 0.05$, and $j = 3$. The initial state is still $\ket{j}$. $\hbar$ is equal to $1$ throughout.}
\label{fig. 2}
\end{figure}

The integrals could again be worked out numerically. Results are shown in Fig.~\ref{fig. 2} for different values of the system-environment coupling strengths, $G$ and $F$, and they are precisely what one would expect them to be if the system Hamiltonian evolution were kept: increasing the weak coupling strength increases the decay rate (Fig.~\ref{fig. 2b}) whereas increasing the strong coupling strength decreases it generally (Fig.~\ref{fig. 2a}). Also, as is evident, a change in the weak coupling strength has no effect on the qualitative behavior of the Zeno/anti-Zeno transitions. A change in the strong coupling strength, however, does have an effect; as before, increasing it causes the transitions to occur at smaller values of $\tau$. Once again, since we obtain these very results even if the system Hamiltonian evolution is kept, we conclude that the system evolution has no practical bearing on any of them.
\null
\vfill
\vfill
\newpage

\section{\label{sec. 3}Discussion}
Our results clearly demonstrate that removing the evolution effected by the system Hamiltonian just before measurements are performed does not change the general effects of strong and weak system-environment couplings.

We expand the previous work done in this area by considering the presence of both strong and weak couplings, and we successfully show not only that independent coupling baths produce independent effects but also that this independence remains intact even after the system evolution is removed, an observation hinting at the deeper independence of independent strong and weak system-environment couplings.

Although we apply our strategy to spin-boson and large spin-boson models only, our work establishes a guide for the implementation of our strategy to other such instances of open quantum systems as well, thereby adding to our knowledge of how multiple system-environment couplings could be treated to study the quantum Zeno and anti-Zeno effects.

Given the ubiquity of open quantum systems, our work is highly relevant, for it could be appended to any investigation of Zeno and anti-Zeno effects in these systems. Just to highlight its scope, we propose that our work could be used to expand investigations of the Unruh effect and Unruh-DeWitt detectors, it could be applied to the studies of nonselective projective measurements, and it could even be employed in analyses of the Zeno and anti-Zeno effects in quantum field theory~\cite{hussain2018decay,facchi2003unstable, majeed2018quantum}. Hence, our work---though simple---has pretty extensive applications.

\begin{acknowledgments}
We would like to extend sincere gratitude to our colleague Hudaiba Soomro for her unstinting support throughout the project.
\end{acknowledgments}


%


\begin{thebibliography}{10}%
\makeatletter
\providecommand \@ifxundefined [1]{%
 \@ifx{#1\undefined}
}%
\providecommand \@ifnum [1]{%
 \ifnum #1\expandafter \@firstoftwo
 \else \expandafter \@secondoftwo
 \fi
}%
\providecommand \@ifx [1]{%
 \ifx #1\expandafter \@firstoftwo
 \else \expandafter \@secondoftwo
 \fi
}%
\providecommand \natexlab [1]{#1}%
\providecommand \enquote  [1]{``#1''}%
\providecommand \bibnamefont  [1]{#1}%
\providecommand \bibfnamefont [1]{#1}%
\providecommand \citenamefont [1]{#1}%
\providecommand \href@noop [0]{\@secondoftwo}%
\providecommand \href [0]{\begingroup \@sanitize@url \@href}%
\providecommand \@href[1]{\@@startlink{#1}\@@href}%
\providecommand \@@href[1]{\endgroup#1\@@endlink}%
\providecommand \@sanitize@url [0]{\catcode `\\12\catcode `\$12\catcode
  `\&12\catcode `\#12\catcode `\^12\catcode `\_12\catcode `\%12\relax}%
\providecommand \@@startlink[1]{}%
\providecommand \@@endlink[0]{}%
\providecommand \url  [0]{\begingroup\@sanitize@url \@url }%
\providecommand \@url [1]{\endgroup\@href {#1}{\urlprefix }}%
\providecommand \urlprefix  [0]{URL }%
\providecommand \Eprint [0]{\href }%
\providecommand \doibase [0]{https://doi.org/}%
\providecommand \selectlanguage [0]{\@gobble}%
\providecommand \bibinfo  [0]{\@secondoftwo}%
\providecommand \bibfield  [0]{\@secondoftwo}%
\providecommand \translation [1]{[#1]}%
\providecommand \BibitemOpen [0]{}%
\providecommand \bibitemStop [0]{}%
\providecommand \bibitemNoStop [0]{.\EOS\space}%
\providecommand \EOS [0]{\spacefactor3000\relax}%
\providecommand \BibitemShut  [1]{\csname bibitem#1\endcsname}%
\let\auto@bib@innerbib\@empty
\bibitem [{\citenamefont {Chaudhry}(2017)}]{chaudhry2017quantum}%
  \BibitemOpen
  \bibfield  {author} {\bibinfo {author} {\bibfnamefont {A.~Z.}\ \bibnamefont
  {Chaudhry}},\ }\bibfield  {title} {\bibinfo {title} {The quantum zeno and
  anti-zeno effects with strong system-environment coupling},\ }\href@noop {}
  {\bibfield  {journal} {\bibinfo  {journal} {Scientific reports}\ }\textbf
  {\bibinfo {volume} {7}},\ \bibinfo {pages} {1} (\bibinfo {year}
  {2017})}\BibitemShut {NoStop}%
\bibitem [{\citenamefont {Leggett}\ \emph {et~al.}(1987)\citenamefont
  {Leggett}, \citenamefont {Chakravarty}, \citenamefont {Dorsey}, \citenamefont
  {Fisher}, \citenamefont {Garg},\ and\ \citenamefont
  {Zwerger}}]{leggett1987dynamics}%
  \BibitemOpen
  \bibfield  {author} {\bibinfo {author} {\bibfnamefont {A.~J.}\ \bibnamefont
  {Leggett}}, \bibinfo {author} {\bibfnamefont {S.}~\bibnamefont
  {Chakravarty}}, \bibinfo {author} {\bibfnamefont {A.~T.}\ \bibnamefont
  {Dorsey}}, \bibinfo {author} {\bibfnamefont {M.~P.}\ \bibnamefont {Fisher}},
  \bibinfo {author} {\bibfnamefont {A.}~\bibnamefont {Garg}},\ and\ \bibinfo
  {author} {\bibfnamefont {W.}~\bibnamefont {Zwerger}},\ }\bibfield  {title}
  {\bibinfo {title} {Dynamics of the dissipative two-state system},\
  }\href@noop {} {\bibfield  {journal} {\bibinfo  {journal} {Reviews of Modern
  Physics}\ }\textbf {\bibinfo {volume} {59}},\ \bibinfo {pages} {1} (\bibinfo
  {year} {1987})}\BibitemShut {NoStop}%
\bibitem [{\citenamefont {Matsuzaki}\ \emph {et~al.}(2010)\citenamefont
  {Matsuzaki}, \citenamefont {Saito}, \citenamefont {Kakuyanagi},\ and\
  \citenamefont {Semba}}]{matsuzaki2010quantum}%
  \BibitemOpen
  \bibfield  {author} {\bibinfo {author} {\bibfnamefont {Y.}~\bibnamefont
  {Matsuzaki}}, \bibinfo {author} {\bibfnamefont {S.}~\bibnamefont {Saito}},
  \bibinfo {author} {\bibfnamefont {K.}~\bibnamefont {Kakuyanagi}},\ and\
  \bibinfo {author} {\bibfnamefont {K.}~\bibnamefont {Semba}},\ }\bibfield
  {title} {\bibinfo {title} {Quantum zeno effect with a superconducting
  qubit},\ }\href@noop {} {\bibfield  {journal} {\bibinfo  {journal} {Physical
  Review B}\ }\textbf {\bibinfo {volume} {82}},\ \bibinfo {pages} {180518}
  (\bibinfo {year} {2010})}\BibitemShut {NoStop}%
\bibitem [{\citenamefont {Chaudhry}\ and\ \citenamefont
  {Gong}(2013)}]{chaudhry2013role}%
  \BibitemOpen
  \bibfield  {author} {\bibinfo {author} {\bibfnamefont {A.~Z.}\ \bibnamefont
  {Chaudhry}}\ and\ \bibinfo {author} {\bibfnamefont {J.}~\bibnamefont
  {Gong}},\ }\bibfield  {title} {\bibinfo {title} {Role of initial
  system-environment correlations: A master equation approach},\ }\href@noop {}
  {\bibfield  {journal} {\bibinfo  {journal} {Physical Review A}\ }\textbf
  {\bibinfo {volume} {88}},\ \bibinfo {pages} {052107} (\bibinfo {year}
  {2013})}\BibitemShut {NoStop}%
\bibitem [{\citenamefont {Silbey}\ and\ \citenamefont
  {Harris}(1984)}]{silbey1984variational}%
  \BibitemOpen
  \bibfield  {author} {\bibinfo {author} {\bibfnamefont {R.}~\bibnamefont
  {Silbey}}\ and\ \bibinfo {author} {\bibfnamefont {R.~A.}\ \bibnamefont
  {Harris}},\ }\bibfield  {title} {\bibinfo {title} {Variational calculation of
  the dynamics of a two level system interacting with a bath},\ }\href@noop {}
  {\bibfield  {journal} {\bibinfo  {journal} {The Journal of chemical physics}\
  }\textbf {\bibinfo {volume} {80}},\ \bibinfo {pages} {2615} (\bibinfo {year}
  {1984})}\BibitemShut {NoStop}%
\bibitem [{\citenamefont {Koshino}\ and\ \citenamefont
  {Shimizu}(2005)}]{koshino2005quantum}%
  \BibitemOpen
  \bibfield  {author} {\bibinfo {author} {\bibfnamefont {K.}~\bibnamefont
  {Koshino}}\ and\ \bibinfo {author} {\bibfnamefont {A.}~\bibnamefont
  {Shimizu}},\ }\bibfield  {title} {\bibinfo {title} {Quantum zeno effect by
  general measurements},\ }\href@noop {} {\bibfield  {journal} {\bibinfo
  {journal} {Physics reports}\ }\textbf {\bibinfo {volume} {412}},\ \bibinfo
  {pages} {191} (\bibinfo {year} {2005})}\BibitemShut {NoStop}%
\bibitem [{\citenamefont {Breuer}\ \emph {et~al.}(2002)\citenamefont {Breuer},
  \citenamefont {Petruccione} \emph {et~al.}}]{breuer2002theory}%
  \BibitemOpen
  \bibfield  {author} {\bibinfo {author} {\bibfnamefont {H.-P.}\ \bibnamefont
  {Breuer}}, \bibinfo {author} {\bibfnamefont {F.}~\bibnamefont {Petruccione}},
  \emph {et~al.},\ }\href@noop {} {\emph {\bibinfo {title} {The theory of open
  quantum systems}}}\ (\bibinfo  {publisher} {Oxford University Press on
  Demand},\ \bibinfo {year} {2002})\BibitemShut {NoStop}%
\bibitem [{\citenamefont {Hussain}\ and\ \citenamefont
  {Ahmed}(2018)}]{hussain2018decay}%
  \BibitemOpen
  \bibfield  {author} {\bibinfo {author} {\bibfnamefont {A.}~\bibnamefont
  {Hussain}}\ and\ \bibinfo {author} {\bibfnamefont {H.}~\bibnamefont
  {Ahmed}},\ }\bibfield  {title} {\bibinfo {title} {Decay of qubits under
  arbitrary space-time trajectories: The zeno \& anti-zeno effects},\
  }\href@noop {} {\bibfield  {journal} {\bibinfo  {journal} {arXiv preprint
  arXiv:1811.09432}\ } (\bibinfo {year} {2018})}\BibitemShut {NoStop}%
\bibitem [{\citenamefont {Facchi}\ and\ \citenamefont
  {Pascazio}(2003)}]{facchi2003unstable}%
  \BibitemOpen
  \bibfield  {author} {\bibinfo {author} {\bibfnamefont {P.}~\bibnamefont
  {Facchi}}\ and\ \bibinfo {author} {\bibfnamefont {S.}~\bibnamefont
  {Pascazio}},\ }\bibfield  {title} {\bibinfo {title} {Unstable systems and
  quantum zeno phenomena in quantum field theory},\ }in\ \href@noop {} {\emph
  {\bibinfo {booktitle} {Fundamental Aspects of Quantum Physics}}}\ (\bibinfo
  {publisher} {World Scientific},\ \bibinfo {year} {2003})\ pp.\ \bibinfo
  {pages} {222--246}\BibitemShut {NoStop}%
\bibitem [{\citenamefont {Majeed}\ and\ \citenamefont
  {Chaudhry}(2018)}]{majeed2018quantum}%
  \BibitemOpen
  \bibfield  {author} {\bibinfo {author} {\bibfnamefont {M.}~\bibnamefont
  {Majeed}}\ and\ \bibinfo {author} {\bibfnamefont {A.~Z.}\ \bibnamefont
  {Chaudhry}},\ }\bibfield  {title} {\bibinfo {title} {The quantum zeno and
  anti-zeno effects with non-selective projective measurements},\ }\href@noop
  {} {\bibfield  {journal} {\bibinfo  {journal} {Scientific reports}\ }\textbf
  {\bibinfo {volume} {8}},\ \bibinfo {pages} {1} (\bibinfo {year}
  {2018})}\BibitemShut {NoStop}%
\end{thebibliography}
\end{document}